\documentclass[11pt,superscriptaddress,showpacs,aps,nofootinbib,preprint]{revtex4}
\pdfoutput=1 %necessary for pdf figures (pdflatex) to JHEP
\usepackage[latin1]{inputenc}
\usepackage{verbatim}
\usepackage{graphicx}
\usepackage{hyperref}
\usepackage{ae,aecompl}
\usepackage{amsmath,amssymb,mathrsfs}
\usepackage[usenames,dvipsnames]{color}

\makeatletter
\@ifundefined{textcolor}{}
{%
 \definecolor{BLACK}{gray}{0}
 \definecolor{WHITE}{gray}{1}
 \definecolor{RED}{rgb}{1,0,0}
 \definecolor{GREEN}{rgb}{0,1,0}
 \definecolor{BLUE}{rgb}{0,0,1}
 \definecolor{CYAN}{cmyk}{1,0,0,0}
 \definecolor{MAGENTA}{cmyk}{0,1,0,0}
 \definecolor{YELLOW}{cmyk}{0,0,1,0}
}

\providecommand{\ZZ}{\mathbb{Z}}
\providecommand{\lag}{\mathscr{L}}
\providecommand{\aver}[1]{\left\langle #1 \right\rangle}
\providecommand{\vsg}{v_\sigma}
\providecommand{\vsl}{v_{\sigma'}}

\providecommand{\Agrav}{A_{\textrm{grav}}}
\providecommand{\Mp}{M_{\textrm{Pl}}}
\providecommand{\vw}{v_{\rm w}}
\providecommand{\eq}[1]{\begin{equation} #1 \end{equation}}
\providecommand{\eqali}[1]{\begin{equation}\begin{aligned} #1
    \end{aligned}\end{equation}}
\providecommand{\ums}[2][1]{\ml{\tfrac{#1}{#2}}} \providecommand{\ml}[1]{\mbox{\large $#1$}}
\providecommand{\tp}{{\mss{\mathsf{T}}}}
\providecommand{\mss}[1]{\mbox{\scriptsize $#1$}}
\providecommand{\xlink}[1]
  {\href{http://arxiv.org/abs/#1}{arXiv:#1}}

\begin{document}

\title{Axion Like Particles and the Inverse Seesaw Mechanism }

\author{C. D. R. Carvajal}
\email{crisdaruiz@gmail.com}
\affiliation{Universidade Federal do ABC, Centro de Ciências Naturais e Humanas,
Av. dos Estados, 5001, 09210-580, Santo André, SP, Brasil}

\author{A. G. Dias}
\email{alex.dias@ufabc.edu.br}
\affiliation{Universidade Federal do ABC, Centro de Ciências Naturais e Humanas,
Av. dos Estados, 5001, 09210-580, Santo André, SP, Brasil}

\author{C.~C.~Nishi}
\email{celso.nishi@ufabc.edu.br}
\affiliation{Maryland Center for Fundamental Physics, 
University of Maryland, College Park, MD 20742, USA}
\affiliation{Universidade Federal do ABC,
Centro de Matemática, Computação e Cognição,
Av. dos Estados, 5001, 09210-580, Santo André, SP, Brasil}

\author{B. L. Sánchez$-$Vega}
\email{brucesanchez@anl.gov}
\affiliation{Argonne National Laboratory, 9700 S. Cass Avenue, Argonne, Illinois 60439, USA}

\begin{abstract}
Light pseudoscalars known as axion like particles (ALPs) may
be behind  physical
phenomena like the Universe transparency to ultra-energetic photons, the soft
$\gamma$-ray excess from the Coma cluster, and the 3.5 keV line. We explore the
connection of these particles with the inverse seesaw (ISS) mechanism for
neutrino mass generation. We propose a very restrictive setting where the
scalar field hosting the ALP is also responsible for generating the ISS
mass scales through its vacuum expectation value on gravity
induced nonrenormalizable operators. A discrete gauge symmetry protects the theory
from the appearance of overly strong gravitational effects and discrete anomaly
cancellation imposes strong constraints on the order of the group.
The anomalous U$(1)$ symmetry leading to the ALP is an extended
lepton number and the
protective discrete symmetry can be always chosen as a subgroup of a combination of the
lepton number and the baryon number.
\end{abstract}
\maketitle

\section{Introduction}

Pseudo Nambu-Goldstone bosons of very low masses, interacting effectively with the
electromagnetic field, are generally predicted in Standard Model (SM) extensions
containing approximate continuous global symmetries which are spontaneously broken.
A distinctive example of this type of particles is the axion,
which arises when the strong CP problem is solved through the
Peccei-Quinn mechanism~\cite{Peccei:1977hh,Weinberg:1977ma,Wilczek:1977pj}.
Generically, any light pseudo Nambu-Goldstone boson whose coupling to
photons is similar to that of the axion has been denoted as an axion like particle
(ALP).
Several experiments are in search of the peculiar effect of photon
$\leftrightarrow$ ALP oscillations and these searches guide
the development of theories containing these
particles~\cite{Jaeckel:2010ni,Ringwald:2012hr,Essig:2013lka}.

The symmetry associated to the ALP is usually taken to be spontaneously broken at
a very high energy scale, and we assume here that this breakdown occurs through
the vacuum expectation value (vev) of a scalar field singlet under the SM
symmetry group.
The ALP decay constant, which controls the feeble ALP couplings to
other SM particles, can be identified to this vev, up to order one coefficients.

In this work we investigate settings where the same scalar singlet hosting the
ALP also gives rise to the mass scales involved in the inverse seesaw (ISS)
mechanism for neutrino mass generation, a well-known mechanism that does not require
too heavy neutral fermions~\cite{iss}.
There are two scales beyond the SM in the ISS mechanism. One of them
is directly related to the lepton number breakdown. In our proposal the usual
lepton number is embedded in an anomalous U(1)$_X$ symmetry associated to the ALP.
Such a symmetry is an accidental one in the sense that it automatically arises from
the imposition of another symmetry considered to be more fundamental.
Breaking of the accidental U(1)$_X$ symmetry is expected from gravitational
interactions through nonrenormalizable operators, which also generate the ALP
mass.

Therefore, we also deal with the problem of stabilizing both the mass scales in the
ISS mechanism and the ALP mass in face of gravitational interactions.
Nonrenormalizable operators that would bring too large mass corrections can be
avoided by assuming discrete $\ZZ_N$ symmetries, which are remnants of
gauge symmetries valid at very high energies~\cite{Krauss:1988zc}.
The choices for the $\ZZ_N$ symmetries are greatly reduced by the conditions
they need to satisfy in order to be free from anomalies~\cite{anomaly.1,anomaly.2}.
For example, there are only a handful of interesting anomaly free discrete gauge 
$\ZZ_N$ symmetries for the MSSM\,\cite{anomaly.1,P6}.

Discrete symmetries have already been used for suppressing dangerous operators
that prevents the solution of the strong CP problem in different models containing the axion~\cite{Lazarides:1985bj,Dine:1992vx,Dias:2002hz,Dias:2002gg,Babu:2002ic,
Carpenter:2009zs,Montero:2011tg,Harigaya:2013vja,Dias:2014osa,Celis:2014jua}.
Such symmetries are shown to be crucial in multi-ALP models,
where very low mass ALPs need effective protection against
gravitational interactions~\cite{Dias:2014osa}.

At the same time that the U(1)$_X$ symmetry breaking scale leads to the correct
mass scales for the ISS mechanism, we look for models where the ALP coupling
with photons and the ALP mass have values that allow the explanation of
three hinted astrophysical phenomena: the anomalous Universe transparency for very
energetic $\gamma$-ray
~\cite{Mirizzi:2007hr,DeAngelis:2007dy,Aharonian:2007wc,Aliu:2008ay,
Essey:2011wv,Horns:2012fx,Simet:2007sa,SanchezConde:2009wu,Meyer:2013pny,rubtsov},
the soft X-ray excess from the Coma
cluster~\cite{lieu96,Cicoli:2012aq,Higaki:2012ar,Conlon:2013txa,Angus:2013sua},
and the X-ray line at 3.5
keV~\cite{Bulbul:2014sua,Boyarsky:2014jta,Higaki:2014zua,Jaeckel:2014qea}.
These phenomena have already motivated the development of
general multi-ALP models, containing an axion dark matter candidate
whose decay constant is associated to the high energy scale entering
in the canonical seesaw mechanism~\cite{Dias:2014osa}.

The first astrophysical hint, the anomalous Universe transparency, follows from
observations of active galactic nuclei
(AGN)~\cite{Aharonian:2007wc,Aliu:2008ay,Essey:2011wv,Horns:2012fx,rubtsov}.
An ALP coupling with the  photon  and with mass within some specific
interval could provide an explanation to this phenomena through $\gamma$-ray
$\leftrightarrow$ ALP oscillations in the magnetic fields of
AGNs, and in the intergalactic
medium~\cite{DeAngelis:2007dy,Simet:2007sa,SanchezConde:2009wu,Meyer:2013pny}
(see, however, Ref.\,\cite{biteau} for a recent analysis of the cosmic transparency 
hint).

Second, the observed excess of soft X-ray coming from the Coma
cluster~\cite{lieu96} could be explained assuming a cosmic ALP background radiation,
originated from the decay of heavy moduli fields and corresponding to a fractional
number of extra neutrinos~\cite{Cicoli:2012aq,Higaki:2012ar}.
The cluster magnetic field would make possible the conversion of the ALP into the
observed X-rays~\cite{Conlon:2013txa,Angus:2013sua}.

At last, the 3.5 keV line has been reported from observations, using the XMM-Newton
satellite data, of the stacked spectrum of galaxy clusters and  in the Perseus
cluster ~\cite{Bulbul:2014sua}, and also in the Andromeda
galaxy~\cite{Boyarsky:2014jta}.
A possible interpretation for this could be the two photon decay of a
dark matter ALP with mass equal to 7.1 keV~\cite{Higaki:2014zua,Jaeckel:2014qea}
(another possibility could be a CP even scalar\,\cite{babu}, or a specific 
majoron\,\cite{majoron}).
Even if a 7.1 keV dark matter does not correspond to an ALP, it could decay into 
ALPs that in turn decay into photons\,\cite{conlon}.
It has to be said that a study with the Chandra data of X-ray of the Milky Way
did not show a conclusive evidence for the 3.5 keV line~\cite{Riemer-Sorensen:2014yda},
and that other interpretations for the 3.5 keV line in terms of some specific
Potassium and Chlorine lines were also suggested~\cite{Jeltema:2014qfa}.
On the other hand it is argued in~\cite{Boyarsky:2014ska} that the
interpretation of dark matter decay as the origin of the 3.5 keV line
is consistent with the XXM-Newton dataset of the Milk Way center.
Although there is some debate on the origin of the 3.5 keV line signal
we shall assume it is due to an ALP decay.

We also motivate our work with the new generation of proposed experiments which
are projected to probe some  regions of the parameter space of the
ALP coupling with photons and its mass. Among
these experiments, we can mention: the ALPS-II~\cite{Bahre:2013ywa}, the helioscope
IAXO~\cite{Armengaud:2014gea}, and the observatories PIXIE~\cite{Kogut:2011xw} and
PRISM~\cite{Andre:2013nfa}. Most of the models we propose here are within the
prospected search range of these experiments. In Figure~\ref{gxm} it is
shown the regions in the parameter space to be tested by these experiments,
as well as the ones allowing for explanation of the hinted astrophysical
phenomena.

The outline of the paper is the following: in Section~\ref{sec:iss} we present the
general setting that relates the physics of an ALP, its astrophysical
motivations and the generation of the ISS scales.
In subsequent Section~\ref{sec:sym}, we analyze the general symmetry properties of
the models and establish necessary conditions for interesting models. Then
in sections \ref{sec:z13m} and \ref{sec:2ALP}  we show, respectively,
models with one and two ALPs.
Finally, We conclude in Section~\ref{sec:conclusion}.

\section{ALP and the inverse seesaw mechanism}
\label{sec:iss}

We start by showing the main elements that need to be considered in
our constructions containing just one complex scalar field whose vev generates the
energy scales  involved in the ISS mechanism, and which are assumed to be
associated with new physics beyond the SM.
Such scales are taken as being proportional to a scalar field vev
times a suppression factor, composed by this vev divided by the Planck scale and
raised to some power.
The complex scalar field hosts an ALP which, through its effective interaction
with the electromagnetic field, is going to provide explanation for astrophysical
phenomena like the soft X-ray excess and the Universe transparency. For this,
the ALP needs to have its mass  protected from dangerous effective operators due to
gravitational interactions. In order to obtain the natural ISS mechanism scales
and the appropriate mass for the ALP, we look for suitable discrete symmetries
over the fields.

In the ISS mechanism~\cite{iss}, two extra sets
of neutral fermionic singlet fields, $N_{iR}$ and $S_{iR}$, $i=1,2,3$, are taken into
account in addition to the SM neutrino fields $\nu_{iL}$.
It is assumed that after spontaneous symmetry breaking, a mass Lagrangian is
generated containing the following terms
\begin{equation}
-\lag \supset \overline{N_{R}}\,m_D\,\nu_{L}+\overline{S_{R}}\,M \,N_{R}^{c}+
\frac{1}{2}\overline{S_{R}}\,\mu\, S_{R}^{c}+\textrm{H.c.}\,,
\label{liss}
\end{equation}
with the $3\times3$ Dirac mass matrices $m_D$, $M$, and the Majorana mass matrix
$\mu$, which without loss of generality can be taken diagonal. The mass matrix
texture arising from Eq. (\ref{liss}), with the basis choice
$[\nu_L\,,N_R^{c}\,,S_R^{c}]$, is
\begin{equation}
{\rm \bf M}=\left[\begin{array}{ccc}
0 & m_D^\tp & 0\\
m_D & 0 & M^\tp\\
0 & M & \mu
\end{array}\right]\, .\label{mtextura}
\end{equation}
\noindent
It was observed by Mohapatra and Valle~\cite{iss} that a mass matrix of the form
in Eq.~\eqref{mtextura} may lead to three active neutrinos with masses
at the sub-eV scale, without invoking very large entries in the matrix ${\rm \bf M}$.
For example, masses at the sub-eV scale for the active neutrinos can be obtained with
the entries of $m_D$, $M$, and $\mu$ of order 10\,GeV, 1\,TeV, and 1\,keV, respectively.
Specifically, the lepton number is only broken by a \textit{small} scale set by
$\mu$, which is the {\it inverse} of what is assumed in the canonical seesaw
mechanism where lepton number is broken by a very large right-handed neutrino scale.
Taking a matrix expansion in powers of $M^{-1}$, block diagonalization of ${\rm \bf M}$
leads to the approximate mass matrix for the three light active neutrinos
\begin{equation}
m_{\nu}\approx m_{D}^{\tp}\, M^{-1}\mu \left(M^{\tp}\right)^{-1}m_D\,,
\label{mnua}
\end{equation}
and a $6\times6$ matrix
\begin{equation}
M_R\approx\left[\begin{array}{cc}
0 & M^{\tp}\\
M & \mu
\end{array}\right]\,,\label{mr}
\end{equation}
related to six neutrinos. These last ones are supposedly heavier than the
active neutrinos,  with masses at the scale of $M$, and are quasi-degenerate
(pseudo-Dirac nature) if the entries of $\mu$ are small compared to the ones in $M$.
If the number of $S_{iR}$ fields were greater than the number of $N_{iR}$ fields,
one or more neutrino states with masses at the $\mu$ scale would arise, and they
could also contribute as dark matter\,\cite{abada} 
(another possibility for keV DM within the ISS mechanism is given in 
Ref.\,\cite{dev:kevDM}).
The mixing between the heavy neutrinos and active neutrinos is
approximately given by $\epsilon=m_D\,M^{-1}$ and unitarity violation effects are
typically of the order $\epsilon^2$. General aspects of the ISS mechanism concerning
the neutrino mixing and violation of unitarity  were developed in
Refs.~\cite{lindner,valle}. Generically, $\epsilon^2$ at the percent level
is not excluded experimentally, but may be within the reach of future experiments
probing lepton flavor violating transitions\,\cite{valle} and direct production
of heavy states at colliders\,\cite{direct}.

The scales involved in $M$ and $\mu$ are supposed to arise from
new physics beyond the SM. In particular, the $\mu$ term in
Eq. \eqref{liss} breaks the lepton number symmetry explicitly.
From the point of view of naturalness it is reasonable
that the nonvanishing entries of $\mu$ be associated with a small
effective energy scale compared to the electroweak scale, $\vw=246$ GeV.
In the limit $\mu\rightarrow 0$  lepton number conservation is restored
increasing the set of symmetries. Thus, the entries of $\mu$
are expected to be small in comparison with the mass scales of
the SM, which contain the lepton number as a global automatic symmetry.

In our approach, the parameters $\mu$ and $M$ are gravity induced and result from 
the very high  vev  of the complex scalar field times suppression factors containing 
the Planck scale, with the parameters in $m_D$ proportional to $\vw$. This contrast 
with proposals where the typical energy scale in $M$ is due to a new theory with 
spontaneous symmetry breaking at the TeV scale  
\cite{Dias:2011sq,Dias:2012xp,Freitas:2014fda}.

The complex scalar field is a singlet under the SM gauge group and has a vev
denoted as $\langle \sigma\rangle=v_\sigma/\sqrt 2$, with
\eq{
	\label{vev.range}
	10^9\,{\rm GeV}\lesssim v_\sigma \lesssim 10^{14}\,{\rm GeV}\,,
}
defining the intermediate scale range. This leads to a photon-ALP
coupling with value required to explain astrophysical phenomena,
with the  ALP detectable by
future experiments~\cite{Dias:2014osa,Ringwald:2014vqa}. We parameterize
the scalar singlet as
\begin{equation}
\sigma(x)=\frac{1}{\sqrt 2}[v_\sigma+\rho(x)]e^{i\frac{a(x)}{v_\sigma}}\,,
\label{sigma}
\end{equation}
were $a(x)$ is the ALP field. The radial field $\rho(x)$ gets a mass
at the scale $v_\sigma$ and we assume it decouples from the low energy
effective theory.
In the models presented here $\sigma$  carries charge of a global U(1)$_X$
chiral symmetry which is explicitly broken by the gravitational interactions in such a
way that, after spontaneous symmetry breaking, the ALP gets a small mass. The
U(1)$_X$ symmetry is taken as accidental meaning that it results from one or more
imposed gauge discrete $\ZZ_N$ symmetries -- not broken by gravitational
interactions -- restricting the main interactions of the neutral fermion fields with
the scalar fields being
\begin{equation}
\lag \supset \overline{N_R}\,y\,\widetilde{H}^\dagger{L}
+\frac{\sigma^{p}}{M_{\textrm{Pl}}^{p-1}}\overline{S_{R}}\,\eta\,N_{R}^{c}
+\frac{1}{2}\frac{\sigma^{q}}{M_{\textrm{Pl}}^{q-1}}\overline{S_R}\,\zeta\,S_{R}^{c}
+\textrm{H.c.}\,,
\label{liss-sh}
\end{equation}
where  $y$, $\eta$ are complex $3\times 3$ matrices, and $\zeta$ is a symmetric
$3\times 3$ matrix. $L_i$ and $H$ are the leptons and Higgs SU(2)$_L$ doublet fields,
respectively, with $\widetilde H=i\tau_2 H^*$.
The complex conjugate field $\sigma^*$ can be equally considered in the third term,
instead of $\sigma$, while we conventionally define the scalar present in the
second term to be $\sigma$. We use the reduced Planck scale
$M_{\textrm{Pl}}= 2.4\times 10^{18}\,\rm GeV$ for the gravitational scale.
The vev of the Higgs doublet field is $\langle H\rangle=(0,\vw/\sqrt 2)^T$.
We will see that U(1)$_X$ is directly related to an extended lepton number and 
thus, in our approach, the smallness of $\mu$ follows from its explicit but small 
breaking due to gravity ($1/\Mp$ suppression) and its spontaneous breaking at the 
scale $v_\sigma$.

\begin{figure}[h]
\center
\includegraphics[scale=0.4]{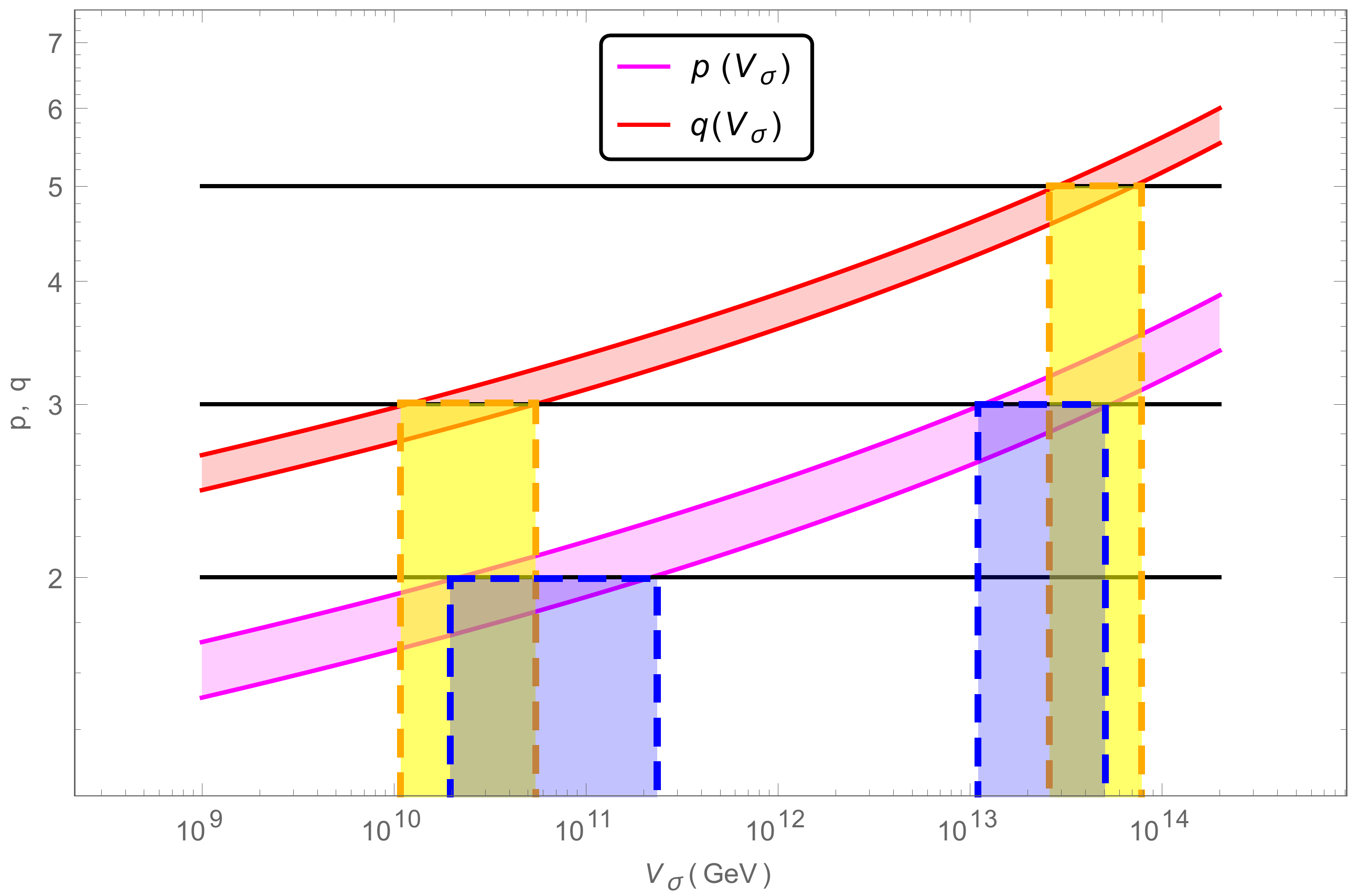}
\protect\caption{$p$ (purple) and $q$ (red) as a function of $v_\sigma$
for $F(p)=\rm 0.1\text{ -- } 10\,TeV$ and $F(q)=\rm 0.1\text{ -- }10\,keV$, with
$F(k)={v_\sigma^k}/{2^{\frac{k}{2}} M_{\textrm{Pl}}^{k-1}}$.
The regions of intersection between yellow and blue bands indicate common values 
for $v_\sigma$, possible for the set of integers $(p,q)=(2,3)$ or $(p,q)=(3,5)$.
}
\label{expo}
\end{figure}

With the vev of the scalar fields in Eq. (\ref{liss-sh}), the effective
Lagrangian in Eq. (\ref{liss}) is obtained with the mass matrices
\begin{eqnarray}
m_D = y\frac{\vw}{\sqrt2}\,,\,\,\,\,\,\,\,\,\,
M = \eta\frac{v_\sigma^p}{2^{\frac{p}{2}} M_{\textrm{Pl}}^{p-1}}\,,\,\,\,\,\,\,\,\,\,
\mu = \zeta\frac{v_\sigma^q}{2^{\frac{q}{2}} M_{\textrm{Pl}}^{q-1}}\,.
\label{mMmu}
\end{eqnarray}
\noindent
$m_D$ is naturally at the 100 GeV scale without requiring the entries
of $y$ to be fine tuned. The nonvanishing entries of
$\eta$ and $\zeta$ are all of order one, under the assumption that the
nonrenormalizable interactions in Eq. (\ref{liss-sh}) are exclusively due to
gravitational interactions, whose universal coupling is $1/M_{\textrm{Pl}}$.
Thus, we look for values of $v_\sigma$ in which the mass scale function
$F(k)=v_\sigma^k/2^{\frac{k}{2}} M_{\textrm{Pl}}^{k-1}$ is assumed to have values
$F(p)=\rm 0.1\text{ -- } 10\,TeV$ and $F(q)=\rm 0.1\text{ -- }10\,keV$,
for $p$ and $q$ integers.
In figure \ref{expo}, curves for $p$ and $q$ are shown as functions of $v_\sigma$.
Within the range in Eq. \eqref{vev.range}, we can see that only
\eq{\label{p.q}
p=2,3 ~\text{ and }~
q=3,4,5\,,
}
can generate the appropriate scales for the ISS mechanism.
Moreover, if only one vev accounts for both $M$ and $\mu$,  it can be
seen that there are only two sets of solutions:
\eqali{
  \label{iss.bands}
(p,q)&=(2,3)& \text{ for }~~ \vsg&\approx 2.2\times 10^{10}\,{\rm GeV}\text{ --- } 5.5\times
10^{10}\,{\rm GeV},\cr
(p,q)&=(3,5)& \text{ for }~~ \vsg&\approx 2.8\times 10^{13}\,{\rm GeV}\text{ --- } 5.5\times
10^{13}\,{\rm GeV}.
}
However, we should keep in mind that the scale for $\mu$ is more flexible than $M$.

We observe that for both sets of values of $(p,\,q)$ in Eq. \eqref{iss.bands} the active
neutrinos masses are independent of the Planck mass,
at leading order, and it happens whenever $q=2p-1$.
This feature would not be possible if, e.g.,
neutrinos masses were generated by type-I seesaw with heavy masses induced by
gravity.
In fact, Eq. (\ref{mnua}) leads, in face of
Eq. (\ref{mMmu}), to
\begin{equation}
m_{\nu}\approx y^{\tp}\, \eta^{-1}\zeta \left(\eta^{\tp}\right)^{-1}y
\frac{\vw^2}{\sqrt2 v_\sigma}\,.
\label{mmac}
\end{equation}
This formula shows that the active neutrinos masses have a
suppression factor $\vw/v_\sigma$ in relation to the electroweak scale, $\vw$.
Such an explanation for having small neutrinos masses
resembles the canonical seesaw mechanism and have been observed in
other models implementing the ISS mechanism~\cite{Dias:2011sq}.

Now we show that besides having its value constrained to be
within a range that allows active neutrinos to have masses at the sub-eV scale,
$v_\sigma$ can also furnish an ALP-photon coupling as required for explaining the
mentioned astrophysical phenomena.
It is supposed here that such ALP-photon coupling arises effectively by means of the
$\sigma$ field interaction with a new colorless fermion, $E$, which carries one unit
of electric charge and is a singlet under SU(2)$_L$ group.
Along with  $\sigma$, $N_{iR}$, and $S_{iR}$, we assume that $E$ also carries charge
of U(1)$_X$. Under this symmetry these fields transform as
\begin{eqnarray}
& & \sigma \rightarrow e^{i \beta }\sigma, \nonumber\\
& & N_{iR} \rightarrow e^{i X_{N} \beta } N_{iR}\,,\,\,\,\,\,\,\,
S_{iR} \rightarrow e^{i X_{S} \beta } S_{Ri},\nonumber\\
& & E_L \rightarrow e^{i X_{E_L} \beta } E_L\,,\,\,\,\,\,\,\,\,\,
E_R \rightarrow e^{i X_{E_R} \beta } E_R,
\label{ftux}
\end{eqnarray}
with charges $X_\psi$, normalized such that $X_\sigma=1$.

The effective Lagrangian describing the interactions of the ALP with photons is
\begin{equation}
\lag_{a\gamma} = \frac{1}{2}\partial_\mu a\partial^\mu a -\frac{1}{2}m_a^2\,a^2
- \frac{g_{a\gamma}}{4}\,a\, F_{\mu\nu} \tilde{F}^{\mu\nu}
\, ,\label{aff}
\end{equation}
where $F_{\mu\nu}$ is the electromagnetic field strength and
$\tilde{F}^{\mu\nu}=\epsilon^{\mu\nu\lambda\rho}F_{\lambda\rho}/2$ its dual.
The ALP-photon coupling constant, $g_{a\gamma}$, is given by
\begin{equation}
g_{a\gamma}=\frac{\alpha}{2\pi}\frac{C_{a\gamma}}{v_\sigma}\,,
\label{gag}
\end{equation}
where $\alpha\approx 1/137$, and the anomaly coefficient $C_{a\gamma}$ depends
on the U(1)$_X$ and electric charges of the fermionic fields $\psi_i$,
$X_{\psi_{L,R}}$ and $C_{\rm em}^{(i)}$,
respectively,  according to
\begin{equation}
C_{a\gamma}=2\sum_{\psi}(X_{{\psi}_L}-X_{{\psi}_R})(C_{\rm em}^{(\psi)})^{2}\,.
\label{cag}
\end{equation}
Throughout this work only the field $E$ is chiral under  U(1)$_X$
-- the left- and right-handed components of $E$ have different U(1)$_X$
charges -- so that Eq. (\ref{cag}) reduces to $C_{a\gamma}=2(X_{E_L}-X_{E_R})$.
The effective coupling in Eq. (\ref{gag}) can be obtained through a
rotation of the fermionic fields $\psi\rightarrow e^{iX_\psi\frac{a(x)}{v_\sigma}}\psi$
which does not leave the integration measure invariant, meaning that the U(1)$_X$
symmetry is anomalous (for details see Ref.~\cite{Kim:1986ax}).
With such a rotation it turns out that the ALP has only derivative couplings
with the fermions. We omit the interactions of the ALP with fermions since these
effects are outside the scope of this work.

A nonzero value for $m_a$ in Eq. (\ref{aff}) must be generated by an explicit
breaking of U(1)$_X$, characterizing the ALP as a pseudo Nambu-Goldstone boson.
We follow the assumption that gra\-vi\-ta\-ti\-o\-nal interactions do not respect any
global continuous symmetry and that U(1)$_X$ is explicitly broken by
nonrenormalizable operators suppressed by $M_{\textrm{Pl}}$.
However, as argued by Krauss and Wilczek~\cite{Krauss:1988zc}, discrete symmetries
like $\ZZ_N$, which are leftover of gauge symmetries, are expected to be
respected by gravitational interactions~\footnote{The argument in
Ref.~\cite{Krauss:1988zc} is essentially that gravitational interactions must
respect local symmetries and also any residual $\ZZ_N$ symmetry left in the
effective theory after spontaneous breaking.}
and, therefore, they can prevent the presence of unwanted nonrenormalizable
operators of lower dimensions. Thus, a $\ZZ_N$ preserving operator of some
high dimension $D$  necessarily breaks U(1)$_X$,
\begin{equation}
\lag\supset \frac{g}{M_{\textrm{Pl}}^{D-4}}\sigma^D +\textrm{H.c.}\,\, ,\,\,\,\,\,
(D>4)
\label{opm}
\end{equation}
where $g=|g|e^{i\delta}$, with $|g|$ of order one; we assume the operator in Eq.
\eqref{opm} is the one with lowest dimension with such a property.
In that case, at leading order, a potential for the ALP is generated
\begin{equation}
V(a)\approx -\frac{|g|}{2}\frac{v_\sigma^D}{(\sqrt2 M_{\textrm{Pl}})^{D-4}}
\cos\left[D\frac{a}{v_\sigma}+\delta\right]\,.
\label{pota}
\end{equation}
This furnishes a mass to the ALP which can be very light for a sufficiently high $D$,
\begin{equation}
m_a\approx |g|^{\frac{1}{2}}D\,\frac{v_\sigma}{\sqrt{2}}\times
\left[\frac{v_\sigma}{\sqrt2M_{\textrm{Pl}}}\right]^{\frac{D}{2}-2}\,.
\label{ma}
\end{equation}

\begin{figure}[h]
\center
\includegraphics[scale=0.45]{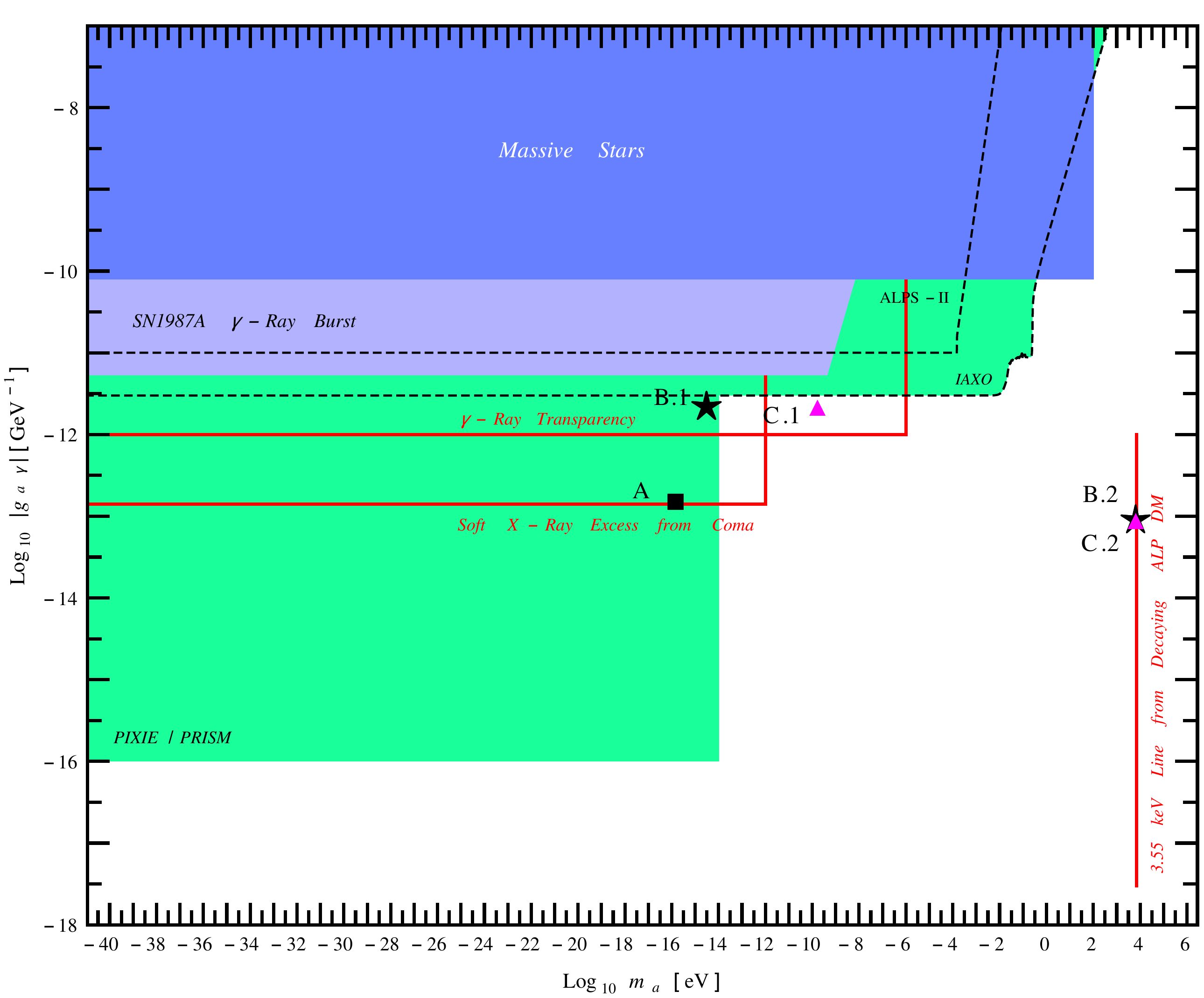}
\protect\caption{Figure adapted from Ref.~\cite{Dias:2014osa}. Values for the ALP
coupling, $g_{a\gamma}$, and mass, $m_a$, required to explain indications of the
anomalous Universe transparency for very energetic
$\gamma$-ray~\cite{Meyer:2013pny},
the soft X-ray excess from the Coma cluster~\cite{Conlon:2013txa,Angus:2013sua},
and the X-ray line at 3.5 keV
~\cite{Higaki:2014zua,Jaeckel:2014qea,Lee:2014xua,Cicoli:2014bfa},
are inside the regions delimited by red lines.
In green are the prospective regions to be reached by the light-shining-through-wall
experiment ALPS-II~\cite{Bahre:2013ywa}, the helioscope IAXO~\cite{Armengaud:2014gea},
and the observatories PIXIE~\cite{Kogut:2011xw} an PRISM~\cite{Andre:2013nfa}.
Also shown are astrophysical limits provided by: emission of ALP from massive
stars representing an anomalous energy loss and shortening their
helium-burning phase so that Cepheids could not
be observed~\cite{Friedland:2012hj},
non-observation of a $\gamma$-ray burst
emitted by the core of the supernova SN 1987A
in coincidence with their neutrinos burst arrival on
Earth~\cite{Grifols:1996id,Brockway:1996yr,new.1987}.
%, and measurements of quasar polarizations~\cite{Payez:2012vf,Payez:2013yxa}.
The benchmark points for the one ALP model of Section\,\ref{sec:z13m} (A, square)
and for the two-ALP models of Section\,\ref{sec:Z8.11} (B.1 and B.2, stars) and
Section\,\ref{sec:Z8.10} (C.1 and C.2, triangles) are also shown.
}
\label{gxm}
\end{figure}

Intervals for the ALP parameters $(g_{a\gamma},\,m_a)$ which can
explain the anomalous Universe transparency for very energetic $\gamma$-ray
~\cite{Mirizzi:2007hr,DeAngelis:2007dy,Aharonian:2007wc,Aliu:2008ay,
Essey:2011wv,Horns:2012fx,Simet:2007sa,SanchezConde:2009wu,Meyer:2013pny},
the soft X-ray excess from the Coma cluster~\cite{Conlon:2013txa,Angus:2013sua},
and the X-ray line at 3.5 keV~\cite{Bulbul:2014sua,Boyarsky:2014jta,Jaeckel:2014qea},
are shown in Figure \ref{gxm}. It can be seen that there is a region where
a set of parameters could explain both the anomalous Universe transparency
and the soft X-ray excess from the Coma cluster.
That region corresponds to $g_{a\gamma}\approx 10^{-11}\text{ --- }
10^{-12}\rm \,GeV^{-1}$, which implies an ALP scale at the range
\eq{
  \label{range:g:transp}
\frac{v_\sigma}{C_{a\gamma}}\approx 10^8\text{ --- }10^9\rm\,GeV
\,,
}
with $m_a\lesssim 10^{-12}\rm eV$.
This requires that the U(1)$_X$ breaking operators in Eq. \eqref{opm} should have
dimensions of at least $D=11$ for $\vsg=10^9\,\rm GeV$, and $D=12$
for $\vsg=10^{10}\,\rm GeV$.

On the other hand, the region of parameters allowed for explaining
the X-ray line at 3.5 keV is disconnected from the previous region.
Thus, if all these hinted phenomena are due to ALPs at least two
different species of them are needed to exist. In order to explain the
3.5\,keV X-ray line through a decay of an ALP with mass 7.1\,keV,
the dimensionality of the operator in Eq. \eqref{opm}  inducing
such a mass depends on the scale $\vsg$.  The values for the ALP scale $\vsg$ and the
operator dimension $D$ inducing the correct mass with coupling in the range
$g_{a\gamma}/C_{a\gamma}\approx 10^{-17}\text{ --- }10^{-12}\rm\,GeV^{-1}$ are
shown in Table \ref{table:7kev}. For simplicity we take $g=1$.
We see that the U(1)$_X$ breaking operator needs to be of dimension 7 or larger.

Large portions of  the ALP parameter space are expected to be probed directly
by new experiments, and are already limited indirectly from astrophysical
observations as shown in Figure \ref{gxm}. Among the direct search experiments
are the light-shining-through-wall experiment ALPS-II~\cite{Bahre:2013ywa},
the helioscope IAXO~\cite{Armengaud:2014gea}, and the observatories
PIXIE~\cite{Kogut:2011xw} an PRISM~\cite{Andre:2013nfa}.
Indirect astrophysical limits excluding portions of the parameter space are
obtained from massive stars~\cite{Friedland:2012hj}, the 1987A 
supernova~\cite{Grifols:1996id,Brockway:1996yr,new.1987}, and quasar 
polarization~\cite{Payez:2012vf,Payez:2013yxa}.
Since the present limit coming from the supernova 1987A\,\cite{new.1987} is 
stronger than the limits coming from quasar polarization, we do not show the latter 
in Figure \ref{gxm}.

\begin{table}[h]
\eq{\nonumber
\begin{array}{|c|c|c|}
\hline
 D & v_\sigma\,[\rm GeV]& g_{a\gamma}/C_{a\gamma}\,[\rm GeV^{-1}]\\
\hline
 7 & 6.04\times 10^8 & 1.92\times 10^{-12}    \\
 8 & 2.44\times 10^{10} & 4.76\times 10^{-14} \\
 9 & 3.43\times 10^{11} & 3.39\times10^{-15} \\
 10 & 2.50\times 10^{12} & 4.64\times10^{-16} \\
 11 & 1.18\times 10^{13} & 9.87\times10^{-17} \\
 12 & 4.07\times 10^{13} & 2.86\times10^{-17} \\
 13 & 1.12\times 10^{14} & 1.03\times10^{-17} \\
\hline
\end{array}
}
\caption{\label{table:7kev}
Necessary values for $\vsg$ to obtain $m_a=7.1\,\rm keV$ in Eq. \eqref{ma} through the
operator $\sigma^D$ in Eq. \eqref{opm}, with $g=1$.
}
\end{table}

Next we present a general analysis of the symmetries and find requirements for an
acceptable model.
After that, we show specific constructions implementing the ISS mechanism
with scales originating from the vev of one or more scalar
fields, which contain ALPs having values of $(g_{a\gamma},\,m_a)$ in the regions
that could explain certain astrophysical phenomena.

%%%%%%%%%%%%%%%%%%%%%%%%%%
\subsection{Symmetries of the model}
\label{sec:sym}

Two symmetries are essential in our construction: the continuous
anomalous U(1)$_X$ symmetry and the stabilizing discrete gauge symmetry $\ZZ_N$.
Here we consider a single factor for simplicity but more factors can be equally
considered.
The continuous U(1)$_X$ symmetry should arise accidentally from the conservation of
$\ZZ_N$ at the intermediate ALP scale. The discrete symmetry, in turn, is assumed
to be a remnant of a continuous gauge symmetry at higher energy scales, possibly
at the GUT scale\,\cite{Krauss:1988zc}.
We will be concerned with the theory at intermediate scales and we will not attempt
to trace the original continuous gauge symmetry as the possibilities are numerous.
The anomalous nature of U(1)$_X$ gives rise to the required ALP-photon coupling
whereas the discrete symmetry $\ZZ_N$ should be anomaly free in the discrete
sense\,\cite{anomaly.1} as required by its local nature.

Let us proceed to determine the generic aspects of the anomalous U(1)$_X$ and the
discrete $\ZZ_N$ symmetries.
We will establish the following: U(1)$_X$ acting on non-SM fields as in
Eq. \eqref{ftux} is an \textit{extension} of the usual lepton number ${\cal L}$ and
$\ZZ_N$ can be chosen as a discrete subgroup of some combination of ${\cal L}$ and the
baryon number ${\cal B}$.

We start with U(1)$_X$ and consider the Yukawa interactions of th SM:
\eq{
  \label{sm:yuk}
-\lag^{\rm SM}_{\rm Yuk}=
\bar{q}_LHd_R+\bar{q}_L\tilde{H}u_R+\bar{L}Hl_R\,,
}
where we omit Yukawa couplings and family indices for simplicity.
The SM fields are denoted as follows: $q_{iL}$ are the left-handed quark
doublets; $u_{iR}$ and $d_{iR}$ are the right-handed quarks singlets;
$l_{iR}$ are the right-handed lepton singlets; with $L_{i}$ and $H$ being
respectively the left-handed lepton and Higgs doublets of Eq.\,\eqref{liss-sh}.

There are three family independent U(1) symmetries in Eq. \eqref{sm:yuk}, coming
from 3 independent constraints on 6 phases associated to 6 types of fields. They can
be identified as hypercharge $Y$, baryon number ${\cal B}$ and lepton number ${\cal L}$.
We adopt the usual assignment that the lepton fields $L_i,l_{iR}$ carry one unit of
lepton number: ${\cal L}=1$.

We now consider the addition of the right-handed neutrino fields $N_R,S_R$,
necessary for the ISS mechanism, and also the complex singlet scalar $\sigma$ whose
vev sets the neutrino mass scales $M,\mu$. These three complex fields contribute to
the Lagrangian in Eq. \eqref{liss-sh}, containing three terms, and no additional U(1)
symmetry appear. These new fields do not carry neither hypercharge nor baryon
number, the latter following from the absence of interactions with quarks. Thus they
carry an extended lepton number. In particular, because of the first term
in Eq. \eqref{liss-sh}, $N_{iR}$ carries the same lepton number as $L_i$.
If we denote by $a,d$ the lepton number of $S_R$ and $\sigma$, respectively,
the last two terms in Eq. \eqref{liss-sh} result in
\eq{
  \label{a.d}
d=(p-\ums{2}q)^{-1}\,,\qquad a=\ums{2}qd\,,
}
where $p\neq q/2$ is required from independency of constraints.
If we exchange $\sigma$ by $\sigma^*$ in the last term in Eq. \eqref{liss-sh}, it is
sufficient to consider negative $q\to -q$ in all equations.
We conventionally adopt positive $p$.

At last, the new vector-like fermion fields $E_L,E_R$ have the same electric charge
as $l_{iR}$ and its hypercharge is defined.
They interact through
\eq{
  \label{lag:E}
-\lag\supset k_i\frac{\sigma^r}{\Mp^r}\bar{L}_iHE_R+
k_E\frac{\sigma^s}{\Mp^{s-1}}\bar{E}_LE_R\,,
}
with small integers $r,s$; note that $s$ cannot be zero to generate an anomalous
symmetry but it should also obey $|s|\le 3$ to induce sufficiently large
masses for $E$, for ALP scales in the range in Eq.  \eqref{vev.range}.
Roughly speaking, the value of $r$ determine the life time of the
charged lepton $E$: the larger the value of $r$, the longer the life time of the
$E$ particle. If $r$ is too large, the first term in Eq. \eqref{lag:E} becomes
negligible, and thus the $E$ exotic lepton will be a stable charged particle which
is cosmologically problematic, unless its mass is $\lesssim$ TeV ~\cite{perl2001}.
Another constraint comes from searches for long-lived charged particles in pp
collisions \cite{Chatrchyan2013}. We will discuss this in more detail below.

The two interaction terms in Eq. \eqref{lag:E} then determine the lepton numbers
of $E_L,E_R$ without affecting the number of symmetries. If we denote the
lepton number of $E_L,E_R$ by $b,c$ respectively, we obtain explicitly
\eq{
  \label{b.c}
b-c=sd\,,\qquad
c=1-rd\,.
}
A negative $s,r$ in Eq. (\ref{b.c}) may account for the simple exchange $\sigma\to
\sigma^*$ in the respective terms.
The final set of U(1) symmetries of the model consists of $Y,{\cal B},{\cal L}$ generated by
charges listed in Table \ref{table:YBL}.
The ${\cal L}$-charges $a,b,c,d$ of fields $S_R,E_L,E_R,\sigma$ were determined by
Eqs. \eqref{a.d} and \eqref{b.c}.
Since ${\cal B}$, and obviously $Y$, are not anomalous with respect to electromagnetism,
the anomalous symmetry U(1)$_X$ can be chosen to be generated by some multiple
of the extended lepton number ${\cal L}$. Specifically, since the anomaly is proportional to
$b-c$, $s$ cannot be zero in view of Eq. \eqref{b.c}.
\begin{table}[h]
\eq{\nonumber
\begin{array}{|c||c|c|c||c||c|c|c||c|c|c||c|}
\hline
 & q_{iL}  & d_{iR} & u_{iR} & H & L_{i}  & l_{iR} & N_{iR} & 	
    S_{iR} & E_{L} & E_{R} & \sigma \cr
\hline
Y & \ums{6} & -\ums{3} & \ums[2]{3} & \ums{2} & -\ums{2} & -1
&  0 & 0 & -1 & -1 & 0\cr
\hline
{\cal B} & \ums{3} & \ums{3} & \ums{3} & 0 & 0 & 0 & 0 & 0 & 0 & 0 & 0\cr
\hline
{\cal L} & 0 & 0 & 0 & 0 & 1 & 1 & 1 & a & b & c & d \cr
\hline
\end{array}
}
\caption{\label{table:YBL} Continuous symmetries of the model. Charges
$a,b,c,d$ are determined by Eqs. \eqref{a.d} and \eqref{b.c}.
}
\end{table}

Concerning  the discrete symmetry $\ZZ_N$, the following anomaly
cancellation conditions should hold from the effective point of
view\,\cite{anomaly.1,anomaly.2}:
\eq{
\label{anomaly:canc}
A_2=A_3=\Agrav=0 \mod N/2\,.
}
where $A_2,A_3,\Agrav$ are the anomaly coefficients associated with
$[\textrm{SU}(2)_L]^2\times \ZZ_N$, $[\textrm{SU}(3)_c]^2\times \ZZ_N$ and
$[\text{gravity}]^2\times\ZZ_N$, respectively.
We ignore the anomaly associated to $[\textrm{U(1)}_Y]^2\times \ZZ_N$ because it does
not furnish useful low energy constraints\,\cite[b]{anomaly.1}

We write the action of $\ZZ_N$ as
\eq{
  \label{def:Z}
\psi_k\to e^{i2\pi Z_k/N}\psi_k\,,
}
with discrete charges $Z_k=Z(\psi_k)=0,1,\cdots,N-1$.
Given that U(1)$_Y$ is anomaly free by construction, and its imposition has no
effect on undesirable operators, we can consider our discrete $\ZZ_N$ to be
a discrete subgroup of the rest of the symmetries in Table
\ref{table:YBL}~\cite{Babu:2002ic,Montero:2011tg}:
\eq{
  \label{Z:BL}
Z=c_1{\cal B}+c_2{\cal L}\,,
}
where $c_i$ should be rational numbers that makes all $Z$ charges integers.
To avoid redundancy, we can adopt $c_1=n_1 3$ and $c_2=n_2\tilde{c}_2$, where
$n_1,n_2=1,\ldots,N-1$ and $\tilde{c}_2$ is the smallest integer that makes all
${\cal L}$-charges integer and coprime. The factor $3$ in $c_1$ appears because
only $3{\cal B}$ is made of integers. If $N$ is not a prime, we also need
to discard values for $n_i$ that makes $c_1{\cal B}$ or $c_2{\cal L}$ to have a
common factor that divides $N$. The latter case implies only a subgroup of
$\ZZ_N$ is realized.

We can now calculate the anomaly coefficients as
\eq{
A_{i}(Z)=c_1A_{i}({\cal B})+c_2A_{i}({\cal L})\,.
}
where
\eqali{
A_{i}({\cal B})&=(\ums[3]{2},0,0)\,,\cr
A_{i}({\cal L})&=(\ums[3]{2},0,-3a+b-c)\,,
}
with $i=2,3,\text{grav}$, respectively.
In special, the gravitational anomaly only depends on ${\cal L}$ and we can write
\eq{
  \label{A.grav}
A_{\rm grav}(Z)=c_2(s-\ums[3]{2}q)d\,,
}
where Eqs. \eqref{a.d} and \eqref{b.c} are used.
We can see the well-known result that ${\cal B- L}$ is anomaly free for $a=0$
and $b=c$, which corresponds to the SM with three right-handed neutrinos;
see e.g. Ref.\,\cite{b-l}.
Therefore, any discrete subgroup of ${\cal B-L}$ will have $A_2$ and $A_3$ automatically
canceled.
However, due to its discrete nature, Eq. \eqref{anomaly:canc}, distinct
combinations of ${\cal B}$ and ${\cal L}$ can be also anomaly free as well.
One can also check, there is no intrinsic discrete symmetry besides subgroups of
combinations of $Y,{\cal B},{\cal L}$; use, e.g., the Smith Normal Form
method~\cite{snf}.

To summarize, we seek SM extensions defined by Eqs. \eqref{liss-sh}
and \eqref{lag:E}, with ALP decay constant $\vsg$, integers $(p,q,r,s)$ and
discrete symmetry $\ZZ_N$ obeying the following restrictions:
\begin{enumerate}
\item One ALP is present that couples to photons and explains one or more
astrophysical phenomena indicated in Fig. \ref{gxm};
\item The correct mass scales for the ISS mechanism are generated by $\vsg$;
\item The ISS mechanism is stabilized by $\ZZ_N$;
\item The mass of the heavy lepton $E$ is larger than the electroweak scale:
$M_E\gtrsim \vw$.
\item There is no discrete anomaly for $\ZZ_N$.
\end{enumerate}
Extensions to more than one ALP should obey analogous conditions.

The conditions for items 1, 2, 4 and 5 have already been discussed.
To summarize conditions 1 and 4, it is necessary to have $0<|s|\le 3$ and the
singlet $\sigma$ should be charged by U(1)$_X$; and $|s|=3$ is possible only if the
ALP scale is high, $\vsg\gtrsim 10^{13}\,\rm GeV$.
The stability of the ISS mechanism, condition 3, requires the following:
$N_R,S_R$
should be charged under $\ZZ_{N}$ to avoid the direct Majorana terms
$\bar{N}_RN_R^c$ and $\bar{S}_RS_R^c$. Moreover, $\ZZ_N$ charges should prevent the
appearance of operators $\sigma^n\bar{S}_{R}S_{R}^c$, $\sigma^n\bar{S}_{R}N_{R}^c$,
 $\sigma^n\bar{N}_{R}N_{R}^c$, $\sigma^n\bar{L}\tilde{H}S_R$ with dimension lower
than the ones inducing the correct ISS scales in condition 2;
and the same applies for operators that replaces $\sigma$ with $\sigma^*$.
Specifically, any operator of the form $\sigma^n\bar{N}_{R}N_{R}^c$ or
$\sigma^{n'}\bar{L}\tilde{H}S_R$ disrupts the zeros in the ISS texture
in Eq. \eqref{mtextura}.
Nevertheless, the mass matrix in Eq. \eqref{mnua} is the leading
contribution as long as $|n'|>|q|-p$, $|n|>2p-|q|$ and
$|n|+|n'|>|q|$; order of magnitude conditions can be extracted
from subleading contributions in the seesaw formula, cf.\,\cite{lindner,dev}.
To guarantee that these contribution are negligible, we require a more strict
condition: $|n|\ge 4$ and $|n'|\ge 3$.
The presence of these dangerous operators can be traced from their
${\cal L}$-charges:
\eq{
  \label{dangerous}
\bar{L}\tilde{H}S_R\sim (q-p)d,\quad
\bar{N}_R^cN_R\sim (2p-q)d\,,
}
where we have conveniently written the charges in terms of the charge of
$\sigma\sim d$. Therefore, the combinations $q-p$ and $2p-q$ control the coupling
of these operators to powers of $\sigma$ and some combinations of $p,q$ can be
readily excluded in the case of one singlet. For example $(p,q)=(2,3)$, is excluded
because it allows both operators in Eq. \eqref{dangerous} to couple to $\sigma^*$.
Generically, it is more interesting to have negative $q$ when $p\neq 0$.

\subsection{Model with one ALP}
\label{sec:z13m}

We focus first on a model which according to our previous considerations
could explain the Universe transparency and the soft X-ray excess
from the Coma cluster. As we pointed out in Figure \ref{gxm}, there
is an overlap in the parameter space and for certain values of $(g_{a\gamma},\, m_{a})$
the same ALP could be responsible for both phenomena.
In addition, to obtain correct order of magnitude parameters for the ISS, we take
the singlet vev to be within the first interval in Eq. \eqref{iss.bands}, corresponding
to the case $(p,q)=(2,3)$.
This choice leads to an ALP-photon coupling constant in Eq. (\ref{gag}) in the range
\eq{
  \frac{g_{a\gamma}}{C_{a\gamma}}
\approx 2.1\times 10^{-14} \text{ --- } 5.3\times 10^{-14}\,\rm GeV^{-1}\,.
}
For a coefficient $C_{a\gamma}$ of order one, the value of $g_{a\gamma}$
would be outside the region required to explain the Universe transparency.
But it would be still possible to explain the soft X-ray excess
from the Coma cluster if $C_{a\gamma}\sim 5$ and the ALP mass is restricted to
$m_{a}\leq10^{-12}$ eV~\cite{Conlon:2013txa,Angus:2013sua}.

We choose the Lagrangian given by Eqs. \eqref{liss-sh} and \eqref{lag:E}, with
$(p,q)=(2,-3)$, $(r,s)=(1,2)$.
For convenience, we write explicitly only the terms that depend on the singlet
$\sigma$:
\begin{eqnarray}
-\mathcal{L} & \supset &
  \eta_{ij}\frac{\sigma^{2}}{M_{\textrm{Pl}}}\overline{S}_{iR}(N_{jR})^{c}
    +\frac{1}{2}\zeta_{ij}\frac{\sigma^{*3}}{M_{\textrm{Pl}}^{2}}\overline{S}_
{iR}(S_{jR})^{c}\nonumber \\
	& + &
k_{i}\frac{\sigma}{M_{\textrm{Pl}}}\overline{L}_{i}HE_{R}+k_{E}\frac{\sigma^{2}}{M_{\textrm{Pl}}}
\overline{E}_{L}E_{R}+\textrm{H.c.}\,.
\label{LYtotalz13}
\end{eqnarray}
The choices above lead to $d=2/7$ in Eq. \eqref{a.d} for the lepton number of
$\sigma$.
As a consequence, only $7{\cal L}$ has all charges integer and the
stability of the ISS mechanism requires either a discrete symmetry $\ZZ_{11}$ or
$\ZZ_{13}$.
Other choices allow operators of the forms
$\bar{L}\tilde{H}S_R\sigma^n,\bar{N}_R^cN_R\sigma^n$ with dimensions that are too
low.
However, because of gravitational anomaly, we choose $\ZZ_{13}$ as it can be seen
below.

The simplest possibility for the anomaly free discrete $\ZZ_{13}$ symmetry is
\eq{
Z=6{\cal B}+7{\cal L}\,,
}
where the coefficients in Eq. \eqref{Z:BL} are chosen as follows:
$c_2=7$ is kept and $c_1=6$ is chosen from the cancellation of mixed
$[\textrm{SU}(2)_L]^2\times\ZZ_{13}$ anomaly.
The explicit anomaly coefficients for $\ZZ_{13}$ are
\eq{
A_2=39/2,~~~A_3=0,~~~\Agrav=13\,,
}
which are all zero modulo $N/2=13/2$. Hence, since the gravitational anomaly only
depends on ${\cal L}$, we can see only $\ZZ_{13}$ is anomaly free and we can discard
$\ZZ_{11}$.
The ISS mechanism is also stable as the lowest order $\ZZ_{13}$ invariant operators
that could disrupt the mechanism are $\sigma^6\bar{N}_R^cN_R$,
$\sigma^5\bar{L}\tilde{H}S_R$.
The explicit charges $Z_i$ for each field can be seen in Table \ref{tabelaz13}.
Note that, for the SM model fields, the $\ZZ_{13}$ charges are
equivalent to $6({\cal B-L})$ modulo 13.
\begin{table}[h]
\eq{\nonumber
\begin{array}{|c||c|c|c||c||c|c|c||c|c|c||c|}
\hline
\ZZ_{13} & q_{iL}  & d_{iR} & u_{iR} & H & L_{i}  & l_{iR} & N_{iR} & 	
    S_{iR} & E_{L} & E_{R} & \sigma \cr
\hline
Z & 2 & 2 & 2 & 0 & 7 & 7 & 7 & -3 & 9 & 5 & 2 \cr
\hline
\end{array}
}
\caption{\label{tabelaz13}
$\ZZ_{13}$ charges $Z_i=(6{\cal B}+7 {\cal L})_i$ in the notation of Eq. \eqref{def:Z}.
}
\end{table}

We can see that the imposition of the $\ZZ_{13}$ symmetry in
Eq.\,(\ref{LYtotalz13}) successfully leads to an accidental U(1)$_{X}$ symmetry,
corresponding to the extended lepton number ${\cal L}$ in Table \ref{table:YBL}, with
charges conveniently rescaled in Table \ref{u1} to give $X_\sigma=1$.
Such a symmetry coincides with the usual lepton number for the SM fields,
but it is anomalous for $E_{L}$, $E_{R}$ fields.
The accidental U(1)$_{X}$ symmetry is only approximately conserved because it is
explicitly broken by higher dimensional $\ZZ_{13}$ invariant operators suppressed
by the Planck scale.
\begin{table}[h]
\eq{\nonumber
\begin{array}{|c||c|c|c||c||c|c|c||c|c|c||c|}
\hline
\textrm{U(1)}_X & q_{iL}  & d_{iR} & u_{iR} & H & L_{i}  & l_{iR} & N_{iR} & 	
    S_{iR} & E_{L} & E_{R} & \sigma \cr
\hline
X & 0 & 0 & 0 & 0 & \ums[7]{2} & \ums[7]{2} & \ums[7]{2} &
  -\ums[3]{2} & \ums[9]{2} & \ums[5]{2} & 1 \cr
\hline
\end{array}
}
\protect\caption{U(1)$_{X}$ charge assignments for the fields in
Eq. \eqref{LYtotalz13}. \label{u1}
}
\end{table}

As the singlet field $\sigma$ acquires a large vev, $\vsg\sim 3\times
10^{10}\rm\,GeV$, the anomalous U(1)$_{X}$ is spontaneously broken,
making its phase field $a(x)$ in Eq. \eqref{sigma} a pseudo Nambu-Goldstone boson.
The singlet vev will also generate the ISS mass parameters
in Eq. \eqref{mMmu} in the correct order of magnitude.
The characteristic shift symmetry for $a(x)$ is broken by $\ZZ_{13}$ invariant
operators of the form in Eq. \eqref{opm}, where $\sigma^{13}$ is the operator of smallest
dimension.
The latter operator gives the dominant contribution in Eq. \eqref{ma} for the ALP mass,
with magnitude
\begin{equation}
m_{a}\simeq 1.58\times10^{-16}{\rm eV}
\,|g|^{\frac{1}{2}}\left(\frac{v_{\sigma}}{3\times10^{10
}\,{\rm GeV}}\right)^{5.5}\,.\label{masxr}
\end{equation}
With the charges in Table \ref{u1} and from Eq.\,(\ref{cag}), we can readily
calculate the anomaly coefficient $C_{a\gamma}=4$ and the ALP-photon coupling,
\begin{equation}
	g_{a\gamma}\simeq1.5\times10^{-13}\left(\frac{3\times10^{10}\,{\rm GeV}}{v_{\sigma}}\right)\,{\rm GeV}^{-1}\,.\label{ga13}
\end{equation}
The benchmark point for this model, named ${\rm {\bf A}}$, is
shown in Figure \ref{gxm}. Although the ALP in this model may explain
the soft X-ray excess from the Coma cluster, it is out of the projected
regions for searches of the ALP-II~\cite{Bahre:2013ywa} and
IAXO~\cite{Armengaud:2014gea} experiments.
But it is inside the region that will be probed by the planned observatories
PIXIE/PRISM~\cite{Kogut:2011xw,Andre:2013nfa}.

Since all U(1)$_X$ and $\ZZ_{13}$ charges are family blind, our model does not
lead to specific predictions for the neutrino flavor structure
(family dependent U(1) symmetries leading to axions has been considered in, e.g., 
ref.\,\cite{khlopov}).
Only the order of magnitude for the absolute mass scale is obtained
through Eq. \eqref{mmac} as
\eq{
  \label{Z13:mnu}
m_{\nu}\approx \big[y^{\tp}\, \eta^{-1}\zeta (\eta^{\tp})^{-1}y\big]
\times 1.4\times 10^{3}\,\rm eV\,,
}
for $\vw=246\,\rm GeV$ and $\vsg=3\times 10^{10}\,\rm GeV$.
The mass matrices in Eq. \eqref{mMmu} for the ISS mechanism acquire quite natural
values as
\eq{
  \label{z13:Mmu}
M= \eta\times 187\,\rm GeV\,,\qquad
\mu= \zeta\times 1.6\,\rm keV\,.
}
Typically we will need the matrix entries of $\eta$ to be larger than one, e.g.,
$\eta\sim 5$, and $y\ll 1$, to evade lepton flavor violation\,\cite{valle} and
direct detection constraints\,\cite{direct} as well as to maintain the validity of
the seesaw formula in Eq. \eqref{mr}.
On the other hand, $\zeta$ can be of order one or smaller. To obtain light neutrino
masses in the sub-eV range, we need small Yukawa coupling, $y\lesssim 0.1$ or
smaller.
The largest eigenvalue of the combination of matrices inside brackets
in Eq. \eqref{Z13:mnu} needs to be at most around $10^{-4}$.

From Table \ref{u1}, we note that the charged lepton $E$ is the
only one that contributes to the coefficient $C_{a\gamma}$ because $E$ has electric
charge different from zero and $X_{E_{L}}\neq X_{E_{R}}$. In this
model, its mass comes from the term $k_{E}\frac{\sigma^{2}}{M_{\textrm{Pl}}}\overline{E}_{L}E_{R}$
in Eq. (\ref{LYtotalz13}). When $\sigma$ gains a vev, $E$ obtains
a mass, $M_E=k_{E}\frac{v_{\sigma}^{2}}{2M_{\textrm{Pl}}}\approx k_E\times 187\,\rm
GeV$.
Therefore, we typically need $k_E$ to be larger than one to avoid the current
lower limit of $574$ GeV on the mass of new charged
leptons\,\cite{Chatrchyan2013}.
To be more specific, the latter limit applies for charged long-lived heavy
lepton with lifetime greater than a few nanoseconds, because these
particles can travel distances comparable to the size of modern detectors
and thus appear to be stable. However, in this model $E$ can decay
into $e_{i}$ and $h^{0}$, with $i=e,\,\mu,\,\tau$ and $h^{0}$
being the Higgs boson with mass of $m_{h^{0}}=125$ GeV. This decay is induced
by the term $k_{i}\frac{\sigma}{M_{\textrm{Pl}}}\overline{L}_{i}HE_{R}$ in
Eq. (\ref{LYtotalz13}). Estimating the lifetime of $E$, $\tau_{E}$,
we find that for $m_E>m_{h^{0}}$, $\tau_{E}$ can be written as
\begin{equation}
	 \tau_{E}\simeq\frac{16\pi}{3\lambda^{2}}\frac{m_{E}^{3}}{\left(m_{E}^{2}-m_{h^{0}}^{2}\right)^{2}}\times6.5822\times10^{-25}\,\textrm{s}.\label{lifetime}
\end{equation}
where we have neglected the masses of the SM leptons, i.e.,
$m_{e},m_{\mu},m_{\tau}\rightarrow0$.
We have also considered that $k_{e}=k_{\mu}=k_{\tau}$. The factor
$\lambda$ in Eq. (\ref{lifetime}) is $\frac{k_{i}v_{\sigma}}{\sqrt{2}M_{\textrm{Pl}}}$.
Taking $k_{e}=k_{\mu}=k_{\tau}=1$, we find that for $m_{E} \gtrsim 250$
GeV, the charged lepton $E$ has a lifetime smaller than $10^{-9}$
s. Therefore, the lower limit of $574$ GeV does not apply and order one values for
$k_E$ are still allowed.

Concerning other possibilities, a few remarks are in order.
Considering the SM augmented by only one singlet scalar $\sigma$ and fermion fields
$N_R,S_R,E_L,E_R$ through Eqs. \eqref{liss-sh} and \eqref{lag:E},
\begin{itemize}
\item there is no other model that contains one ALP capable of explaining the
transparency of the Universe or the soft X-ray excess in the Coma cluster;
\item it is also not possible to find models featuring an ALP with mass $m_a\sim
7.1\,\rm keV$, which can explain the 3.55\,keV X-ray line through the decay of the
ALP into two photons~\cite{Higaki:2014zua,Jaeckel:2014qea}.
\end{itemize}
Other choices for the powers $(p,q)$, such as $(p,q)=(3,\pm5)$,
and for the discrete symmetry $\ZZ_N$ do not comply with one or more of the
restrictions explained in the end of Section\,\ref{sec:sym}: (i) generation of
correct mass scales for the ISS mechanism and (ii) stabilization of the ISS
mechanism and (iii) cancellation of discrete anomalies.
Many possibilities are excluded by (ii) because they allow low-dimensional
operators $\sigma^n$ to couple to $\bar{N}_R^cN_R$ or $\bar{L}\tilde{H}S_R$.
Further restriction comes from the gravitational anomaly cancellation, (iii), and
only the $\ZZ_{13}$ symmetry model survives.

%%%%%%%%%%%%%%%%%%%%%%%%%%%%%%
\subsection{Models with two ALPs}
\label{sec:2ALP}

Here we extend the previous setting and seek models featuring
two ALPs that can explain the excess of X-ray photons in the 3.5 keV line, in
addition to the transparency of the Universe for ultra energetic gamma rays and the
soft X-ray excess from the Coma cluster.
As we have previously discussed and shown in Figure \ref{gxm}, at least two ALPs
are necessary
to explain these three phenomena.
Hence, in addition to the singlet $\sigma$, we introduce another SM singlet
$\sigma'$ which will host a second ALP $a'$. Now, two energy scales,
$\vsg=\sqrt{2}\aver{\sigma}$ and $\vsl=\sqrt{2}\aver{\sigma'}$,
will govern the physics of these ALPs.
Since the ALP $a'$ should also couple to photons, the singlet $\sigma'$
should be charged under another anomalous symmetry U(1)$_{X'}$, which follows
accidentally from a second $\ZZ_{N'}$ symmetry.

Let us choose $a$ to be the ALP of 7.1\,keV mass that explains the 3.55\,keV X-ray
line. The possible values for $\vsg$ and the $\ZZ_N$ symmetry that are needed
can be seen in Table \ref{table:7kev} for $|g|=1$.
The possible values for $\vsl$ are then restricted by Eq. \eqref{range:g:transp},
$\vsl\approx 10^9\,\rm GeV$, and $\sigma'$ should be protected by a discrete
symmetry $\ZZ_{N'}$ with $N'\ge 11$.
Note that the two ALP scales do not mix in our models,
following the ones proposed in Ref~\cite{Dias:2014osa}.
Models where only $\sigma$ (or $\sigma'$) couples to $N_R,S_R$ are excluded
from the considerations of the previous section.
We need that both $\vsg$ and $\vsl$ generate the ISS mass
scales.
However, we were unable to find a plausible model that could satisfy all conditions
listed in the end of Section\,\ref{sec:sym}.
Thus we present in the following, two models that satisfy almost all criteria.

The general Lagrangian we will consider is composed of the usual Yukawa interactions
in Eqs. \eqref{sm:yuk} and \eqref{liss-sh}, with terms depending on the scalar singlets
modified to
\eqali{
-\lag & \supset
\frac{\sigma^p{\sigma'}^{p'}}{\Mp^{p+p'-1}}\overline{S_{R}}\eta N^c_{R}
+\frac{\sigma^q{\sigma'}^{q'}}{\Mp^{q+q'-1}}\overline{S_{R}}\zeta S^c_{R}
\\
 &~~+\ k_{i}\frac{\sigma^r{\sigma'}^{r'}}{\Mp^{r+r'}}\overline{L_{i}}HE_{R}
+k_E\frac{\sigma^s{\sigma'}^{s'}}{\Mp^{s+s'-1}}\overline{E_{L}}E_{R}
+\textrm{H.c.}\, .
\label{lag:2ALP}
}
Many restrictions on the integers $(p,q,r,s)$ discussed in Section~\ref{sec:sym} and
in the beginning of Section~\ref{sec:iss} are now valid for the sum of unprimed and
primed variables. For example, the restriction in Eq. \eqref{p.q} should be
now adapted to $p+p'=2,3$ and $|q|+|q'|=3,4,5$, where we conventionally take $p,p'$
to be positive.
Likewise, condition 4 in the end of Section~\ref{sec:sym}, for low ALP scales, is now
$|s|+|s'|\le 2$, which leads to $s=\pm 1$ and $s'=\pm 1$.
We also see that the number of symmetries are consistent: there is one more
field $\sigma'$ for the same number of constraints but we need one more anomalous
symmetry. Given that the fields beyond the SM only couple to leptons, we can still
consider $X$ and $X'$ proportional to two extended lepton numbers
${\cal L}$ and ${\cal L}'$.
Additionally, all formulas for $(p,q,r,s)$ in Section~\ref{sec:iss} still apply
considering that $\sigma$ is only charged under ${\cal L}$ while $\sigma'$ is only charged
under ${\cal L}'$.
Thus the same formulas applies for the primed $(p',q',r',s')$ as well, depending
now on ${\cal L}'$-charges $(a',b',c',d')$ of $S_R,E_L,E_R,\sigma'$, respectively.

\subsubsection{Model I}
\label{sec:Z8.11}

The first model gives up the cancellation of the gravitational anomaly for one of
the discrete symmetries $\ZZ_{N}$ or $\ZZ_{N'}$.
The anomaly can be easily canceled by the addition of one or more fermions that are
singlets of the SM but do not contribute to the phenomena discussed in this paper.
The model also gives rise to a small scale for $\mu$.

We consider the interaction terms for the singlet fields to be
\eqali{
-\lag & \supset
\eta_{ij}\frac{\sigma^2}{\Mp}\overline{S_{Ri}}N^c_{Rj}
+\zeta_{ij}\frac{{\sigma'}^{*3}}{\Mp^{2}}\overline{S_{Ri}}S^c_{Rj}
\\
 &~~+\ k_{i}\frac{\sigma{\sigma'}}{\Mp^2}\overline{L_{i}}HE_{R}
+k_E\frac{\sigma{\sigma'}}{\Mp}\overline{E_{L}}E_{R}
+\textrm{H.c.}\,.
\label{2ALP:Z8.11}
}
The Lagrangian has the form in Eq. \eqref{lag:2ALP} with $(p,q)=(2,0)$ and
$(p',q')=(0,-3)$, whereas $(r,s)=(r',s')=(1,1)$.
We choose $\vsg\approx 2.44\times 10^{10}\,\rm GeV$ to accommodate the correct
scale for $M$ and generate the ALP mass of $m_a=7.1\,\rm keV$, which should be
protected by a $\ZZ_8$ symmetry.
We protect $\sigma'$ with a symmetry $\ZZ_{11}$ so that the whole model has a
symmetry $\ZZ_8\times\ZZ_{11}$.

Let us proceed to find the symmetry $\ZZ_8\times\ZZ_{11}$.
Eq. \eqref{a.d} determines $(d,a)=(1/2,0)$ and $(d',a')=(2/3,-1)$.
Therefore $c_2=2$ and $c_2'=3$ makes all charges of $c_2{\cal L}$ and $c_2'{\cal L}'$ integers
and we can calculate the gravitational anomaly from Eq. \eqref{A.grav}:
\eq{
\Agrav(c_2{\cal L})=1,\quad
\Agrav(c_2'{\cal L}')=11\,.\quad
}
It is clear that the gravitational anomaly for $c_2{\cal L}$ does not cancel for any
$\ZZ_N$, except $\ZZ_2$. Note that we can not use $c_2=2\times 4$ to cancel the
gravitational anomaly because $8{\cal L}$ only generates $\ZZ_2$.
Therefore, we assume such a gravitational anomaly is
canceled by additional fermion fields and we adopt $\ZZ_8$ generated by
$Z=6{\cal B}+2{\cal L}$ (we could have adopted $Z=6{\cal B}-6{\cal L}$ as well).
This choice cancels the anomaly of $A_2(Z)$.
Analogously, we choose the $\ZZ_{11}$ generator as $Z'=-3{\cal B}+3{\cal L}'$.
We show the explicit charges in Table \ref{tab:Z11Z8}.
The charges for the anomalous U(1)$_X$ and U(1)$_{X'}$ are presented in
Table \ref{XX':charges}.
\begin{table}[h]
\eq{\nonumber
\begin{array}{|c||c|c|c||c||c|c|c||c|c|c||c|c|}
\hline
 & q_{iL}  & d_{iR} & u_{iR} & H & L_{i}  & l_{iR} & N_{iR} & 	
    S_{iR} & E_{L} & E_{R} & \sigma & \sigma'\cr
\hline
\ZZ_8 & 2 & 2 & 2 & 0 & 2 & 2 & 2 & 0 & 2 & 1 & 1 & 0 \cr
\hline
\ZZ_{11} & -1 & -1 & -1 & 0 & 3 & 3 & 3 & -3 & 3 & 1 & 0 & 2 \cr
\hline
\end{array}
}
\caption{\label{tab:Z11Z8}
$\ZZ_{8}\times\ZZ_{11}$ charges in the notation in Eq. \eqref{def:Z}.
}
\end{table}

\begin{table}[h]
\eq{\nonumber
\begin{array}{|c||c|c|c||c||c|c|c||c|c|c||c|c|}
\hline
 & q_{iL}  & d_{iR} & u_{iR} & H & L_{i}  & l_{iR} & N_{iR} & 	
    S_{iR} & E_{L} & E_{R} & \sigma & \sigma'\cr
\hline
X & 0 & 0 & 0 & 0 & 2 & 2 & 2 & 0 & 2 & 1 & 1 & 0 \cr
\hline
X' & 0 & 0 & 0 & 0 & \ums[3]{2} & \ums[3]{2} & \ums[3]{2} & -\ums[3]{2} &
      \ums[3]{2} & \ums{2} & 0 & 1 \cr
\hline
\end{array}
}
\caption{Charges for U(1)$_X$ and U(1)$_{X'}$.}
\label{XX':charges}
\end{table}

This model yields $C_{a\gamma}=C_{a'\gamma}=2$, cf. Eq. \eqref{cag}, which
leads to the desired ALP-photon couplings
\eqali{
g_{a\gamma}=\frac{\alpha}{2\pi\vsg}C_{a\gamma}\approx9.52\times10^{-14}
\,\textnormal{GeV}^{-1}\,,\cr
g_{a'\gamma}=\frac{\alpha}{2\pi\vsl}C_{a'\gamma}\approx2.32\times10^{-12}
\,\textnormal{GeV}^{-1}\,,
}
for $\vsg\approx 2.44\times10^{10}$\,GeV and $\vsl\approx 10^9$\,GeV.
The ALP masses are given by
\eqali{
m_{a}\approx |g|\times 7.1\,{\rm keV},~~
m_{a'}\approx |g'|\times 3.41\times 10^{-15}\,\rm eV\,.
}
Benchmark points for ALPs $a$ and $a'$ are marked as B.1 and B.2 in
Figure \ref{gxm} for $|g|=|g'|=1$.

The induced neutrino mass matrices have magnitude
\eq{
M= \eta\times 124\,\text{GeV},\quad
\mu= \zeta\times 0.061\,\text{eV}\,,
}
which leads to the light neutrino mass matrix
\eq{
  \label{Z8.11:mnu}
m_{\nu}= \big[y^{\tp}\, \eta^{-1}\zeta (\eta^{\tp})^{-1}y\big]
\times 0.12\,\rm eV\,.
}
Analogously to the model of Section~\ref{sec:z13m}, we typically need $\eta$ to have
entries with magnitude larger than one while the Yukawa coefficients need to be
smaller than one, so that $\epsilon=m_DM^{-1}$ has small entries. The matrix
$\zeta$ can have entries of order one or smaller but we can see the scale generated
by $\sigma'$ is smaller than the one generated in Eq. \eqref{z13:Mmu}.
The ISS mechanism is stable as the new operators of lowest order are
$N_R^2(\sigma^4{\sigma'}^{3})^*$ and $\bar{L}\tilde{H}S_R\sigma^2{\sigma'}^3$.

\subsubsection{Model II}
\label{sec:Z8.10}

In the second model featuring two ALPs, we require the cancellation of all
anomalies, including gravitational anomalies, but we relax the conditions for
stability of the ISS mechanism. Because of the former, we can only find a
symmetry $\ZZ_8\times \ZZ_{10}$, so that the ALP $a'$ is heavier than the previous
model and can account for the $\gamma$-ray transparency problem but not the soft
X-ray from the Coma cluster.

The model Lagrangian involving $\sigma,\sigma'$ is
\eqali{
-\lag & \supset
\eta_{ij}\frac{\sigma^2}{\Mp}\overline{S_{Ri}}N^c_{Rj}
+\zeta_{ij}\frac{\sigma^{*2}{\sigma'}^{*}}{\Mp^{2}}\overline{S_{Ri}}S^c_{Rj}
\\
 &~~+\ k_{i}\frac{\sigma{\sigma'}}{\Mp^2}\overline{L_{i}}HE_{R}
+k_E\frac{\sigma{\sigma'}}{\Mp}\overline{E_{L}}E_{R}
+\textrm{H.c.}\, .
\label{2ALP:Z8.10}
}
The Lagrangian has the form in Eq. \eqref{lag:2ALP} with
$(p,q)=(2,-2)$ and $(p',q')=(0,-1)$, whereas $(r,s)=(r',s')=(1,1)$.
As in the previous model, we choose $\vsg\approx 2.44\times 10^{10}\,\rm GeV$ and
$\sigma$ is protected by $\ZZ_8$.
The symmetry $\ZZ_{10}$ then protects $\sigma'$.

After performing the calculations of Section~\ref{sec:sym} for this case,
we choose $Z=3{\cal L}-3{\cal B}$ and $Z'={\cal L}'+9{\cal B}$
as generators of  $\ZZ_8$ and $\ZZ_{10}$, respectively;
they are given in Table \ref{charges:Z8.10}.
One can check that these charges are anomaly free.
The anomalous symmetries U(1)$_X$ and U(1)$_{X'}$ can be obtained from
the same table by eliminating the baryon number contributions and
rescaling the $X'$ charge of $\sigma'$ to unity. The extended lepton
numbers ${\cal L}$ and ${\cal L}'$ can be extracted in an analogous manner.
They give $C_{a\gamma}=C_{a'\gamma}=2$.
The ALP--photon couplings are the same as for model I, 
\eq{
g_{a\gamma}\approx 9.52\times10^{-14}\,\textnormal{GeV}^{-1}\,, \quad
% g_{a'\gamma}\approx 4.65\times10^{-12}\,\textnormal{GeV}^{-1}\,,
g_{a'\gamma}\approx 2.32\times10^{-12}\,\textnormal{GeV}^{-1}\,,
}
for $\vsg\approx 2.44\times10^{10}$\,GeV and $\vsl\approx 10^9$\,GeV.
The ALP masses are given by
\eqali{
m_{a}\approx |g|\times 7.1\,{\rm keV},~~
m_{a'}\approx |g'|\times 1.81\times 10^{-10}\,\rm eV\,.
}
We can see $m_{a'}$ is too large to explain the soft X-ray excess in the Coma cluster.
Benchmark ALP photon couplings and masses can seen on Fig. \ref{gxm} marked as
C.1 and C.2.
\begin{table}[h]
\eq{\nonumber
\begin{array}{|c||c|c|c||c||c|c|c||c|c|c||c|c|}
\hline
 & q_{iL}  & d_{iR} & u_{iR} & H & L_{i}  & l_{iR} & N_{iR} & 	
    S_{iR} & E_{L} & E_{R} & \sigma & \sigma'\cr
\hline
\ZZ_8 & -1 & -1 & -1 & 0 & 3 & 3 & 3 & -1 & 3 & 2 & 1 & 0 \cr
\hline
\ZZ_{10} & 3 & 3 & 3 & 0 & 1 & 1 & 1 & -1 & 1 & -1 & 0 & 2 \cr
\hline
\end{array}
}
\caption{\label{charges:Z8.10}
$\ZZ_{8}\times\ZZ_{10}$ charges in the notation in Eq. \eqref{def:Z}.
}
\end{table}

Finally, the neutrino mass matrices have magnitude
\eq{
M= \eta\times 124\,\text{GeV},\quad
\mu= \zeta\times 36.5\,\text{eV}\,,
}
which leads to the light neutrino mass matrix
\eq{
  \label{Z8.11:mnu}
m_{\nu}= \big[y^{\tp}\, \eta^{-1}\zeta (\eta^{\tp})^{-1}y\big]
\times 72\,\rm eV\,.
}
In this case, we have a more natural scale for $\mu$ compared to the model of
Section\,\ref{sec:Z8.11} but the general considerations for $\eta$ are the same.
Concerning the stability of the ISS mechanism, we can see the lowest order
operators that disrupt the texture in Eq. \eqref{mtextura} are
$N_R^2\sigma^2{\sigma'}^*$ and $\bar{L}\tilde{H}S_R\sigma^{*2}\sigma'$.
They lead respectively to mass parameters of magnitude
\eq{
\mu_N\sim\frac{\vsg^2\vsl}{2^{3/2}\Mp^2}\approx 36.5\,\text{eV},~~
m_{DS}\sim\frac{\vw\vsg^2\vsl}{2^{4/2}\Mp^3}\approx 2.6\times 10^{-15}\,\text{eV}.
}
These mass matrices contribute to entries (3,1) and (2,2) in Eq. \eqref{mtextura} and
contribute to subleading terms in the light neutrino mass matrix in Eq. \eqref{mnua}
as\,\cite{lindner,dev}
\eqali{
\delta m_\nu =&
-[m_D^\tp M^{-1}m_{DS}+m_{DS}^\tp M^{\tp -1}m_D]-m_{DS}^{\tp}M^{\tp -1}\mu_N M^{-1}m_{DS}\\
& +m_{D}^{\tp}M^{-1}\mu M^{\tp -1} \mu_N M^{-1}m_{DS} + m_{DS}^{\tp} M^{\tp -1} \mu_{N} M^{-1}\mu M^{\tp -1} m_{D}
 \\
& - m_{D}^\tp M^{-1}\mu M^{\tp -1}\mu_N M^{-1} \mu M^{\tp -1} m_{D}+O(M^{-5},\mu^{3},\mu_{N}^{2}).
}
We can see the contribution of $m_{DS}$ is negligible. The mass parameter $\mu_N$
is of the order of $\mu$ but it also contributes negligibly to the light neutrino
mass matrix, even if one-loop corrections are taken into account\,\cite{dev}.

%%%%%%%%%%%%%%%%%%%%%%%%%%%%%%%
\section{Conclusions}
\label{sec:conclusion}

We have generically studied the construction of models where one ALP results
from a scalar singlet carrying an anomalous extension of the lepton number
${\cal L}$ of the SM.
The ALP successfully accounts for some intriguing astrophysical phenomena such as
the soft X-ray excess in the Coma cluster, and at the same time, the singlet vev
furnishes the correct mass scales that implements the ISS mechanism for
neutrino mass generation through gravity induced nonrenormalizable terms.
Moreover, the approximate nature of anomalous ${\cal L}$-number, the ALP mass
and the ISS mechanism are protected from additional gravity induced terms
through a gauge discrete symmetry.
The additional beyond SM fields are minimal: three families of
right-handed neutrino fields $N_{iR},S_{iR}$, one heavy singlet lepton $E$
and one singlet scalar $\sigma$ are added.

By requiring the stability of the ISS mechanism and the cancellation of the
discrete gauge anomalies, only one model survives, and the discrete symmetry needs
to be a $\ZZ_{13}$ subgroup of a combination of ${\cal L}$ and the baryon
number $B$, the simplest being $6{\cal B}+7{\cal L}$.

Simple extensions to models with two ALPs can be constructed by adding solely one
more singlet scalar. In this case, two ALPs can solve more astrophysical phenomena
with distinct features.
We have been unable to find a model capable of explaining the three astrophysical
phenomena and, at the same time, satisfying all the conditions in
Section\,\ref{sec:sym}.
Therefore, two models are presented by relaxing some of the conditions.
The first model can explain all the astrophysical phenomena, but the gravitational
anomaly for one $\ZZ_{N}$ factor can not be canceled within the field content,
and additional fermionic fields are required. The second model does not present
discrete anomaly but it can not explain the soft X-ray excess in the Coma cluster.
In both cases, the 7.1\,keV ALP may be an appreciable component of dark
matter as well\,\cite{Dias:2014osa,alp.dm}.

In summary, we have proposed a very restrictive and economical setting to extend
the SM and explain notable astrophysical phenomena together with natural neutrino
mass generation through the ISS mechanism.
The restrictiveness of the setting allows only one model with one ALP and models
with two ALPs are also largely restricted.
More possibilities emerge if we allow the presence of more than one heavy
charged lepton and, in particular, three copies of them can lead to easier
cancellation of the gravitational anomaly.

%%%%%%%%%%%%%%%%%%%%%%%%%%%%%%%
\section*{Erratum}

{\small
The two models containing two ALPs in Sec.\,2.C.1
% Sec.\,II.C 
% of Ref.\,\cite{paper} 
lead to too 
light non-SM charged leptons $E$ of only a few GeV.
(The single ALP model of Sec.\,2.C.2 does not suffer from this 
problem.)

This problem can be amended without significant modifications by considering two 
heavy vector-like fermions $E$ and $E'$ instead of one.
Instead of the last two terms of
% Eq.\,(2.40), (2.41) and (2.47) of \cite{paper}, 
Eq.\,(40), (41) and (47),
we should consider the four terms	
\eqali{
-\lag & \supset
 k_{i}\frac{\sigma{\sigma'}}{\Mp^{2}}\overline{L_{i}}HE_{R}
+ k'_{i}\frac{\sigma{\sigma'}}{\Mp^{2}}\overline{L_{i}}HE'_{R}
\cr&~~+
k_E\sigma\overline{E_{L}}E_{R}+k_E'\sigma'\overline{E'_{L}}E'_{R}
+\textrm{H.c.}\,,
\label{lag:2ALP}
}
where we already fixed the $r,r'$ powers without affecting any formula.
One should use $s=s'=1$ if needed. Now the heavy leptons $E,E'$ have intermediate 
scale masses of order $10^{9\div 10}\rm GeV$.
The charge of the discrete symmetries are only modified for $E_{L,R}$ and 
$E'_{L,R}$ in a predictable manner; see table below. 
Appropriate normalization of these charges leads to the PQ charges $X,X'$ and the 
extended lepton numbers $L,L'$.
Note that $U(1)_X$ [$U(1)_{X'}$] is vectorial for $E'$ ($E$) and that the first two 
terms of the Lagrangian above implies $E_R$ and $E_R'$ have equal charges 
($X,X',Z,Z',L,L'$).
The rest of the formulas and the phenomenological consequences remain unchanged.
\begin{table}[h]
\eq{\nonumber
\begin{array}{|c||c|c|c|c|}
\hline
 & E_{L} & E_{R} & E'_{L} & E'_{R} \cr
\hline
\ZZ_8 & 2 & 1 & 1 & 1 \cr
\hline
\ZZ_{11} & 1 & 1 & 3 & 1 \cr
\hline
\end{array}
% }
\hspace{3em}
% \eq{\nonumber
\begin{array}{|c||c|c|c|c|}
\hline
 & E_{L} & E_{R} & E'_{L} & E'_{R}\cr
\hline
\ZZ_8 & 3 & 2 & 2 & 2 \cr
\hline
\ZZ_{10} & -1 & -1 & 1 & -1 \cr
\hline
\end{array}
}
\caption{
Corrected charges for model I (left) and model II (right).
}
\end{table}

Additionally, the lowest order operator coupling $\bar{L}$-$S_R$ for model II is 
$\bar{L}\tilde{H}S_R\sigma^4\sigma'$ instead of 
$\bar{L}\tilde{H}S_R\sigma^{2}{\sigma'}^*$.
This implies $m_{DS}$ is negligible in Eq.\,(52).
}

\bigskip

%%%%%%%%%%%%%%%%%%%%%%%%%%%%%%%
{\it Acknowledgements.}
The authors thank Andreas Ringwald for very helpful suggestions.
This research was  supported by  Conselho Nacional de Desenvolvimento Cient\'{\i}fico e Tecnol\'ogico (CNPq)  (A.G.D and C.D.R.C), 
and by the grants 2013/22079-8 (A.G.D and C.C.N) and 2013/26371-5 (C.C.N) of
Funda\c{c}\~{a}o de Amparo \`{a} Pesquisa do Estado de S\~ao Paulo (FAPESP).
C.C.N. also thanks the 
\textit{Maryland Center for Fundamental Physics} for its hospitality.
B.L.S.V  thanks Coordena\c{c}\~ao Aperfei\c{c}oamento de
Pessoal de N\'{\i}vel Superior (CAPES), for the financial support under Contract
No. 2264-13-7 and the Argonne National Laboratory for its hospitality.

%%%%%%%%%%%%%%%%%%%%%%%%%%%%%%

\end{document}